\documentclass{article}
\usepackage{setspace}
\usepackage[utf8]{inputenc}
\usepackage{amsmath}
\usepackage{slashed}
\usepackage{amssymb}
\usepackage{bbold}
\usepackage{halloweenmath}
\usepackage{graphicx}
\usepackage{authblk}
\usepackage{caption}
\graphicspath{ {./images/} }
\usepackage{comment}
\usepackage{tikz}
\usepackage{geometry}
\usepackage{hyperref}
\hypersetup{ 
colorlinks=true,
linkcolor=blue,
filecolor=magenta,
urlcolor=blue,
citecolor=blue,
pdfpagemode=FullScreen
}
\usepackage[numbers,sort&compress]{natbib}

\usepackage[normalem]{ulem}
\usepackage{enumerate}
\usepackage[shortlabels]{enumitem}

\title{Scalar Charges in a Self-Dual Background}
\author[1]{Seth Grable}
\author[1]{Jamison Barcelona}

\affil[1]{Department of Physics, University of Colorado Boulder, Colorado 80309, USA}
\date{July, 2026}

\begin{document}
\begin{spacing}{1.5}
\maketitle

\begin{abstract}
In this work, we present a \(3+1d\) scalar QED model in a constant self-dual magnetic field configuration. We provide exact closed form analytic calculations of the partition function, the \(\beta\)-function, and the propagators. To our knowledge, we present the first closed-form finite expression for the matter field propagators in a self-dual background, which is made accessible by working entirely in the Landau level basis. This theory serves as a toy model for charged particles in parallel electric and magnetic fields, with natural extensions to studies of \(3+1d\) chiral magnetic effects and pulsar physics.   
\end{abstract}

In natural units, a covariantly constant self-dual field configuration models constant parallel electric and magnetic fields \(E\) and \(B\), with \(E=B\). Physical situations featuring this gauge configuration include: pair production showers and coherent high-energy gamma-ray emission in pulsars, the chiral magnetic effect in relativistic heavy-ion collisions, and magnetoresistance in Dirac and Weyl semimetals \cite{philippov2022pulsar, fukushima2008chiral, yan2017topological}. Self-dual configurations where \(E=B\) also give rise to non-perturbative mathematical properties in field theories such as Yang-Mills instanton solutions in \(3+1\)d Euclidean space \cite{belavin1975pseudoparticle}.

In this work, we present an analytically solvable model for charged scalar fields in a constant self-dual magnetic background. All methods used are non-perturbative, placing no analytic restrictions on the background field strength. This field theory is similar to the scalar version of an Euler-Heisenberg Lagrangian \cite{dunne2005heisenberg, heisenberg2006consequences} with a constant self-dual background configuration. The Euler-Heisenberg Lagrangian is a 1930s formulation of quantum field theory that predates Feynman diagram perturbation techniques. Since its time, the Euler-Heisenberg methodology has been employed by Schwinger to study pair production out of the vacuum \cite{schwinger1951gauge}, and various authors such as Savvidy, Nielsen, Olesen, and Leutwyler have used similar methodologies in modeling the Yang-Mills vacuum non-perturbatively as chromomagnetic flux vortices generated by a constant chromomagnetic background
\cite{savvidy1977infrared, nielsen1978approximate,nielsen1978unstable, leutwyler1981constant}. Modern uses of the Euler-Heisenberg Lagrangian include calculating the birefringence of the vacuum around magnetars \cite{valluri2026vacuum}. Magnetars generate fields strong enough to cross the Schwinger limit, spontaneously creating electron-positron pairs that polarize the vacuum, giving it crystal-like and birefringent properties — effects that are now in the early stages of observation \cite{stewart2025vacuum}.

Further, recent work on charged particles in magnetic fields can be found in \cite{ghosh2024anisotropic, jaber2023scalar, miransky2015quantum, ghosh2025neutrino, ghosh2025neutrino2,shovkovy2025review}. In these latter works, the Landau level structure of the magnetic model is emphasized; however, the authors work in the canonical momentum basis when calculating properties of the self-energies, emission rates, and propagators. In contrast, we present the field theory calculations in the Landau level basis, which gives rise to finite, exact, and closed-form analytic calculations of the partition function and the matter field propagators. The propagators of the Euler-Heisenberg Lagrangian have been known for decades in integral form \cite{dunne2005heisenberg}, however, the integrals are typically divergent or intractable, which we believe is due to working in the momentum basis, in which the theory is not diagonalizable. To our knowledge, no closed finite functional forms of the propagators have previously been presented.

\subsection{Paper Overview}

\paragraph{}
In this paper, we present three calculations that form the backbone of any quantum theory: 1) an eigenvalue analysis of the Hamiltonian, 2) a path integral calculation of the partition function and running coupling, and 3) a calculation of the propagators. All work done is in closed analytic form, using Schrödinger picture techniques in a quantum field theory format. Despite being formulated as a field theory, all calculations are non-perturbative.

We start in section \ref{Sec 1} by giving a brief overview of how quadratic field theories, with arbitrary spatial couplings like \(x^2\phi^2(x)\) can be treated as a Schr\"odinger like eigenvalue problem at the level of the Wick-rotated classical Lagrangian. In \ref{Sec 2} we provide a Euclidean path integral for scalar charges in a magnetic field. This is analogous to a scalar Euler-Heisenberg Lagrangian in a constant self-dual gauge. In section \ref{sec 3} we study the Lorentz signature Hamiltonian. Applying a similarity transformation and first and second quantization conditions, we show that the Hamiltonian is positive definite, Hermitian, and gapped. It follows that the associated field theory is also gapped and unitary. The gapped nature of the Hamiltonian comes from the fact that the first quantization (which takes place at the classical level) is that of a quantum harmonic oscillator. Incidentally, the second quantization is also a quantum harmonic oscillator, so the system gives a gapped spectrum of field excitations, which are an artifact of localized interactions around a constant background. These excitations, however, act as non-interacting quanta. In section \ref{sec 4}, we perform a Schr\"odinger like eigenvalue problem on the classical Wick-rotated Lagrangian. In section \ref{sec 5}, with the use of zeta-function regularization and heat-kernel techniques, we give a non-perturbative calculation of the zero temperature partition function, and the \(\beta\)-function of the running coupling. Finally, in section \ref{Sec 6} we calculate the propagators of the theory and show that field interactions give Gaussian decay across spatial and temporal separations.

\section{Computational Background}\label{Sec 1}
\paragraph{} 

The Euclidean space path integral formulation of quadratic scalar theories evaluates as \cite{laine2016basics,grable2022theremal}
\begin{equation}
    \int \mathcal{D}\phi e^{-\int_x \phi \hat{\theta} \phi} = e^{-\frac{1}{2}\ln\det(\hat{\theta})},
\end{equation}
where \(\mathcal{L}_E = \phi \hat{\theta} \phi\) is the Euclidean Lagrangian density, \(\det(\hat{\theta}) = \prod_i \lambda_i\), and \(\lambda_i\) is an eigenvalue of \(\hat{\theta}\). We have expressed \(\mathcal{L}_E\) in terms of some differential operator \(\hat{\theta}\) to emphasize that the evaluation of quadratic path integrals reduces to an eigenvalue problem similar to a time-independent Schr\"odinger equation:
\begin{equation}
    \hat{\theta}\phi_i = \lambda_i \phi_i
\end{equation}
where \(\phi_i\) is an eigenstate, or an \textit{eigenfield} of \(\hat{\theta}\). However, unlike solving a Schr\"odinger problem, at this level, the eigenvalue quantization of the fields is a quantization of the Wick rotated Lagrangian density, and \(\lambda_i\) has units of energy squared.

 Similarly to the Schr\"odinger picture, eigenfield configurations of some \(\hat{\theta}\) are built from basis states for translations under which \(\hat{\theta}\) is invariant.
That is, If an operator \(\hat{\theta }\) is invariant under a transformation generated by a Hermitian operator \(\hat{\alpha }\) via the unitary \(U = e^{i \xi \hat{\alpha}}\), then \([\hat{\theta}, U] = 0\) and \([\hat{\theta}, \hat{\alpha}] = 0\). Since \(\hat{\theta }\) and \(\hat{\alpha}\) commute, they share a simultaneous basis of eigenstates. Therefore, the fields of the theory can be expanded in the basis of these generator eigenstates, reflecting the underlying symmetry of \(\hat{\theta}\). So long as the action is a scalar quantity, it is invariant under Lorentz transformations in the Minkowski metric, or similarly, it is invariant to translations and global rotations in the Euclidean metric. However, states of various quadratic operators are not necessarily invariant to translations plus rotations in the Euclidean metric. When states of the Lagrangian density operator break a certain symmetry while that symmetry is maintained by the action this is a generalized version of spontaneous symmetry breaking. However, there is no original symmetry of the states that is spontaneously broken. For this reason we refer to this as \textit{state symmetry breaking.}

 For example, considering a Klein-Gordon equation \([-\partial_\mu^2 +m^2]\phi= E^2 \phi\) gives \(\phi = e^{\pm i\sqrt{E^2 -m^2}x_\mu}\). With \(E^2= p^2 +m^2\), then \(\phi=e^{\pm i p x}\) giving \(\phi =U_p\), where \(U_p\) are momentum eigenstates, matching the fact that the Klein-Gordon operator \(\hat{\theta} = -(\partial_\mu)^2 +m^2\) is translationally invariant, i.e. diagonalizable in the momentum basis. The system's basis of plane-wave eigenfields is translation-invariant, which yields a gap between the vacuum and the first excited state, and a continuous ungapped spectrum thereafter. With this, the second quantization represents freely propagating field quanta of a continuous energy spectrum.

A novel artifact of state symmetry breaking in a field theory is that fields may now couple directly to spatial terms. For example we could consider a harmonic oscillator form of \(\hat{\theta} = -\partial_\mu^2 + \frac{m^4}{4} x_i^2\) which yields a Lagrangian density of
\begin{equation}\label{baby L}
    \mathcal{L}_E = \phi[-\partial^2_\mu + \frac{m^4}{4} x_i^2]\phi
\end{equation}
with a dynamic mass term of \(m^2(x)=\frac{m^4}{4} x_i^2\), where \(x_i\) are the spatial terms.
While spatially dependent terms are absent in theories with Lorentz-invariant states, they result in solvable field theories for dynamic systems such as particles in a magnetic field, or toy models for systems with dynamic mass and confining effects. 
While the states of these theories break Lorentz invariance, the action \(S\) or \(Z\) more generally retains Lorentz invariance in the Minkowski metric or translation plus global rotation invariance in the Euclidean metric. In our specific model the quadratic operator of the theory, and thus the states of the theory are rotationally invariant in the (0-1) and (2-3) planes. This theory then generates bound orbits around a constant magnetic field. The operators of the model further transform covariantly under Lorentz boost in the Minkowski metric as it is built from covariant derivatives and field strength tensors. The spatial coupling to the fields generated from the magnetic field gives a discrete quantization of the classical Lagrangian in the quantum harmonic oscillator basis. The discrete nature of the theory is what gives rise to a gapped spectrum of bound states beyond the ground state.

\section{Scalar QED in a Constant Background}\label{Sec 2}
\paragraph{}
 The starting point of our theory is Euclidean-space path integral for a minimally coupled scalar QED model in \(3+1d\) as
\begin{equation}\label{Z 2}
    Z= \int\mathcal{D}\phi\mathcal{D}\phi^* e^{-\int_x \frac{1}{4e^2}F_{\mu\nu}^2 + (D_\mu\phi)^*(D_\mu\phi)},
\end{equation}
where all gauge fields are set to a constant classical background configuration. We consider only the dynamics of the matter fields under the influence of a constant magnetic background represented by the gauge field \(A_\mu(x) = -\frac{1}{2} F_{\mu\nu}x_\nu\) \cite{grable2025vanishing, savvidy2023stability}. Our choice of background is a self-dual configuration, with field strength tensor 
\begin{equation}
    F_{\mu\nu} =\begin{pmatrix}
        0&-B&0&0\\B&0&0&0\\0&0&0&-B\\0&0&B&0
    \end{pmatrix},
\end{equation}
and the covariant derivative as 
\begin{equation}
    D_\mu = \partial_\mu + \frac{i}{2} F_{\mu\nu}x_\nu.
\end{equation}
After integration by parts, the matter field action is
\begin{equation}\label{S}
   S_{\text{matter}} = \int_x \phi^*\big(-D_\mu^2\big)\phi.
\end{equation}
where 
\begin{equation}\label{theta}
    -D_\mu^2=-\partial^2_\mu + iB( x_1\partial_0 -x_0\partial_1) + iB( x_3\partial_2 -x_2\partial_3) +\frac{B^2x_\mu^2}{4}.
\end{equation}
The Euclidean-space Lagrangian of this theory has the structure of a Hamiltonian describing two independent rotations and four oscillations. The actual Hamiltonian of the theory shares a similar eigenstructure to \(-D_\mu^2\), which is explored in section \ref{sec 3}. The Lagrangian has a global \(U(1)\) symmetry under the generator \(\phi \to e^{i \alpha}\phi\) and therefore has the associated conserved current of 
\begin{equation}
    J_\mu = i\Big[\phi(D_\mu\phi)^* - \phi^*(D_\mu\phi) \Big].
\end{equation}
Further, the action \eqref{S} is rotationally invariant in the 0-1 and 2-3 planes, so the quadratic operator commutes with generators of rotation in the \(0-1\) and \(2-3\) planes. As a result of this (shown in section \ref{sec 4}), the eigenbasis of \(-D^2_\mu\) consists of angular momentum eigenstates, and the classical Lagrangian is quantized into discrete modes given by two independent Landau level spectra. As shown in section \ref{sec 3}, the discrete nature of the Lagrangian is analogously found in the first quantization of the Hamiltonian, which takes place at the classical level, resulting in a gapped spectrum. As explored in section \ref{sec 4}, solving this field theory starts with treating the classical Lagrangian as a Schrodinger picture eigenvalue problem, and from here the path integral of the theory can be computed non-perturbatively using zeta-function regularization \cite{hawking1977zeta}.

\section{The Matter Field Hamiltonian}\label{sec 3}
\paragraph{}
In this section, we perform first and second quantizations of the matter field Hamiltonian. To do this, we work in the Schr\"odinger picture and diagonalize the spatially dependent operators by expanding the fields in the eigenbasis of the spatial operators. The resulting Hamiltonian is then expressed as an infinite set of harmonic oscillators, while still working in the Schr\"odinger picture. This procedure shows the theory is gapped, and gives the energy levels of the excitations. We note that this is not an exhaustive analysis of the Hamiltonian. 

To obtain the matter field Hamiltonian contribution of our theory, which we refer to as the Hamiltonian throughout this section, we start by reversing the Wick rotation on the time coordinate by letting \(\tau = -it\). Using spacetime components of \((t,x,y,z)\). This gives the Minkowski space matter field action as
\begin{equation}
    -i S = -i\int_x \big[\partial_t\phi^* \partial_t\phi -\nabla \phi^* \nabla \phi -B \phi^*(x\partial_t\phi + t\partial_x\phi) -i B\phi^*(z\partial_y\phi-y\partial_z\phi) - \frac{B^2}{4}(x_i^2-t^2)\phi^*\phi\big].
\end{equation}
The canonical momenta of the Lagrangian density are \(P^* = \partial_t \phi\) and \(P = \partial_t \phi^*-Bx\phi^*\). The Legendre transform \(\mathcal{H}= P^*\partial_t \phi^* + P \partial_t \phi -\mathcal{L}\) of the Lagrangian density is then 
\begin{equation}\label{baby ham}
    \mathcal{H} = |\partial_t\phi|^2 + |\nabla\phi|^2 + B\phi^* t\partial_x \phi +i B\phi^*(z\partial_y\phi-y\partial_z\phi) + \frac{B^2}{4}(x_i^2-t^2)|\phi|^2.
\end{equation}
Strictly speaking, equation \eqref{baby ham} is not the Hamiltonian density, as it is not yet expressed in terms of canonical momenta.
However, our aim is to promote \(\mathcal{H}\) to the quantum operator \(\hat{\mathcal{H}}\) using the canonical variables for \(\Pi\) and \(\phi\). With this the quantum Hamiltonian density \(\hat{\mathcal{H}}\) is 
\begin{equation}
    \Hat{\mathcal{H}} = |\hat{\Pi}|^2 + |\nabla\hat{\phi}|^2 + B\hat{\phi}^* t\partial_x \hat{\phi} +i B\hat{\phi}^*(z\partial_y-y\partial_z)\hat{\phi} + \frac{B^2}{4}(x_i^2-t^2)|\hat{\phi}|^2.
\end{equation}
Similar to the classic quantization of the Klein-Gordon model, our strategy is to first find the eigenspectrum of all spatial operators acting on \(\hat{\phi}\). To do this we start by focusing on all \(x\)-dependent operators in \(\Hat{\mathcal{H}}\). An apparent instability arises from the negative \(-t^2\) term, however there is also a coupling \(t\partial_x\) that must be considered. To understand the mixing between \(x\) and \(t\), consider the \(x\)-operators in \(\Hat{\mathcal{H}}\). Integrating by parts gives
\begin{equation}
    \Hat{\mathcal{H}}_x = \hat{\phi}^{*}\Big( -\partial^2_x + Bt \partial_x + \frac{B^2}{4}(x^2 -t^2)\Big)\hat{\phi}
\end{equation}
Next, we perform a similarity transformation on \(\Hat{\phi}\) with the substitution \(\Hat{\phi}' = e^{i \Lambda x} \hat{\phi}\). Then 
\begin{equation}\label{H x}
    \Hat{\mathcal{H}}_x = \hat{\phi}^{*}\Big( -\partial^2_x + (B t - 2 i \Lambda)\partial_x + \Lambda(i B t+ \Lambda) + \frac{B^2}{4}(x^2 -t^2)\Big)\hat{\phi}
\end{equation}
Setting \(\Lambda = -\frac{iBt}{2} \) the linear derivative term vanishes, as does the negative time dependence, giving
\begin{equation}
    \Hat{\mathcal{H}}_x = \hat{\phi}^{*}\Big( -\partial^2_x + \frac{B^2}{4}x^2\Big)\hat{\phi}.
\end{equation}
The similarity transformation does not affect the path integral measure of \(\mathcal{D\phi^*}\mathcal{D\phi}\).
The complete similarity-transformed quantum Hamiltonian density is then
\begin{equation}
   \Hat{\mathcal{H}} = |\hat{\Pi}|^2 + |\nabla\hat{\phi}|^2  +i B\hat{\phi}^*(z\partial_y-y\partial_z)\hat{\phi} + \frac{B^2}{4}x_i^2|\hat{\phi}|^2.
\end{equation}
The \(x\)-dependent eigenvalue equation is analogous to a quantum harmonic oscillator, and the \(y\) and \(z\) operators give a Landau-level spectrum arising from a combination of two harmonic oscillators. We refer to the eigenfunctions of the spatial operators as \textit{eigenfields}. The key feature of this theory is the spatial operators acting on the eigenfields are not diagonalizable in the momentum basis. As a result, the modes of our field are not freely propagating plane wave solutions, but instead form localized spectra analogous to those generated by a laser trap \cite{phillips1998nobel}. A more complete analysis of the eigenfield basis will be given in section \ref{sec 4} in context of the Euclidean-space Lagrangian. We can expand the fields in the quantum harmonic oscillator eigenbases of the spatial operators to give
\begin{equation}
    \phi(x_i,t) = \frac{1}{\sqrt{C_{nm}}}\sum^\infty_{n,m=0} \phi(t)H_n(x)\psi_{m}(z,y)a_{nm}
\end{equation}
where \(H(x)_n \) and \(\psi(y,z)_{m}\) are the eigenfields of the spatial operators, \(a_{nm}\) are the expansion coefficients, and \(C_{nm}\) is a combined normalization coefficient for \(H(x)_n\) and \(\psi(z,y)_{m}\). As \(H(x)_n\) are Hermite polynomials and \(\psi(z,y)_{m}\) are eigenstates of angular momentum in the \(y-z\) plane, these eigenfields form an orthonormal basis with the appropriate \(C_{nm}\) constant, as shown in section \ref{sec 4}. Letting \(\Tilde{\Pi}_{nm} = \partial_t \phi(t)a_{nm}\) and \(\Tilde{\phi}_{nm} = \phi(t)a_{nm}\), the Hamiltonian becomes
\begin{equation}
     \int d^3 x\Hat{\mathcal{H}} =\sum^\infty_{n,m=0} \bigg(|\hat{\Tilde{\Pi}}|_{nm}^2 + B\Big(n + 2m + \frac{3}{2}\Big)|\Hat{\Tilde{\phi}}|_{nm}^2\bigg).
\end{equation}
By letting \(\omega^2_{nm} = B\Big(n + 2m + \frac{3}{2}\Big)\) the Hamiltonian takes the familiar form of
\begin{equation}
    \Hat{H}= \int d^3 x\Hat{\mathcal{H}} =\sum^\infty_{n,m=0} \Big(|\hat{\Tilde{\Pi}}|_{nm}^2 + \omega^2_{nm}|\Hat{\Tilde{\phi}}_{nm}|^2\Big).
\end{equation}
The real and imaginary contributions of \(|\hat{\Tilde{\Pi}}|\) and \(|\hat{\Tilde{\phi}}|\) can be separated to give each \(m,n\) mode of \(\Hat{H}\) as
\begin{equation}
    \hat{H}_{mn} = (\Hat{\Tilde{\Pi}}^2_{\text{Re}} + \Hat{\Tilde{\Pi}}^2_{\text{Im}})_{mn} + \omega_{mn}^2( \Hat{\Tilde{\phi}}^2_{\text{\text{Re}}} + \Hat{\Tilde{\phi}}^2_{\text{Im}})_{mn}.
\end{equation}
As the real and imaginary components of \(\Hat{\Tilde{\Pi}}\) are orthogonal, we can treat them as independent momentum operators, and likewise for \(\hat{\Tilde{\phi}}\). It follows that \(\Hat{H}_{nm}\) can be decomposed into a double set of harmonic oscillators, giving 
\begin{equation}
    \Hat{H}  =\sum^\infty_{n,m=0} \omega_{nm}(a^{\dagger}_{mn} a_{mn} + b^{\dagger}_{mn}b^{}_{mn}+ 1).
\end{equation}
Thus, the field Hamiltonian is a set of quantum harmonic oscillators governing the spatial field components within a set of quantum harmonic oscillators that generate the field quanta. Because at the second level of quantization, the theory is quadratic in \(\Pi\) and \(\phi\), it has the structure of a free theory at the level of field excitations. The first quantization generates finite volume and energy bound states with a gapped spectrum. However, these gapped excitations propagate as non-interacting quanta.

The sum over \(\omega_{nm} \times 1\) generates an infinite constant associated with the zero-point energy. This constant can be regulated with zeta functions, but we will discard it, as it does not affect the physical spectrum. 
 With this, the Hamiltonian gives a positive-definite gapped spectrum. The first quantization (the first ``quantum'' harmonic oscillator) of our theory gives standing wave solutions to the field configurations where higher \(n\) and \(m\) correspond to higher energy states of the field configuration. The second quantization (or second quantum harmonic oscillator) gives the number of field excitations. The first excited mode for \(m,n=0\) is given by \(a^{\dagger}_{00}|0\rangle = (\sqrt{3B/2})|1\rangle\), and the gap between the vacuum and the first excited state is \(E_{\text{Gap}}= \sqrt{3B/2}\) under the normalization condition \(\langle 0 |0\rangle =1\). As both quantizations of the Hamiltonian are simple quantum harmonic oscillators, it follows that the theory is local, unitary, positive-definite, and it generates finite volume and energy bound states via the first quantization which transform covariantly under Lorentz boosts.

\section{The Classical Lagrangian Eigenvalue problem}\label{sec 4}
\paragraph{}
Returning to the field theory setup, our Euclidean space path integral is defined as
\begin{equation}\label{Z 2}
    Z= \int\mathcal{D}\phi\mathcal{D}\phi^* e^{-\int_x \frac{1}{4e^2}F_{\mu\nu}^2 + \phi^*[-D^2_\mu]\phi}.
\end{equation}
In this section, we perform an eigenvalue analysis of \(-D^2_\mu\), which is the quadratic operator \(\hat{\theta}\) acting on \(\phi\). After finding the spectrum, the partition function is expressed as \(Z=e^{-\ln\det(-D_\mu^2)}\), as shown in section \ref{sec 5}. A similar eigenvalue analysis by Savvidy appears in the appendix of \cite{savvidy2023stability}, in the context of a one-loop self-dual Yang-Mills model, and this approach broadly dates back to Leutwyler \cite{leutwyler1981constant}.
To find the eigenbasis of \(-D^2_\mu\) we write the eigenvalue problem of \(-D^2_\mu\phi = \lambda \phi\) giving
\begin{equation}\label{theta}
   \Big( -\partial^2_\mu + iB( x_1\partial_0 -x_0\partial_1) + iB( x_3\partial_2 -x_2\partial_3) +\frac{B^2x_\mu^2}{4}\Big)\phi = \lambda\phi.
\end{equation}
This structure is analogous to a time-independent \(4d\) Schrödinger equation for a double Landau level problem, implying that the eigenstates can be constructed by a linear combination of creation and annihilation operators. To start with, we introduce the operators \(c^{\dagger}_\mu = (-\partial_\mu + B x_\mu/2)\) and \(c_\mu = (\partial_\mu + B x_\mu/2)\). Generally  \(c^{\dagger}_\mu\) and \(c_\mu\) generate a ladder spectrum. This can be seen from the commutator relation
\begin{equation}
    [c_\mu, c^{\dagger}_\mu] = B
\end{equation}
where we have used the commutator \([\partial_\mu,x_\mu]=1\), which is an alternative form of the canonical quantum commutator \(i[\hat{p}_\mu,\hat{x}_\mu] =1\). 
We then define the linear combinations \(a^{\dagger}_{\mu\nu} = c^{\dagger}_\mu+ ic^\dagger_\nu\) and  \(a_{\mu\nu} = c_\mu - ic_\nu\) as first noted in \cite{leutwyler1981constant}. A direct calculation gives 
 \begin{equation}
      -D^2_\mu = a^{\dagger}_{01}a_{01} + a^{\dagger}_{23}a_{23} + 2B, 
 \end{equation}
and \(-D^2_\mu\) produces an eigenspectrum of
\begin{equation}\label{spec}
\lambda_{n,l} = 2B(n+ l +1).
\end{equation}
 To find the ground state of the system, we solve the differential equation
\begin{equation}
   (a_{01}+a_{23})\psi_{00} = 0 
\end{equation}
 giving the series of decoupled equations \((\partial_\mu + \frac{B x_\mu}{2})\psi_{00} =0 \) 
to give \(\psi_{00}=e^{-\frac{B}{4}x^2_\mu}\). Then the eigenfields of the theory are given by
\begin{equation}
    \psi_{n,l} = (a^{\dagger}_{01})^n(a^{\dagger}_{23})^l\psi_{00}.
\end{equation}
With \((a^{\dagger}_{\mu\nu})^n = (c^{\dagger}_\mu+ ic^\dagger_\nu)^n\) a recursive relation can be derived for the \(n^{\text{th}}\) power of the creation operators \cite{savvidy2023stability}, giving 
\begin{equation}
    (c^{\dagger}_0+ ic^\dagger_1)^n e^{-\frac{B}{4}(x^2_0 + x^2_1)} = B^n(x_0 +ix_1)^n e^{-\frac{B}{4}(x^2_0 + x^2_1)}
\end{equation}
Thus 
\begin{equation}
    \psi_{n,l} = (a^{\dagger}_{01})^n(a^{\dagger}_{23})^l\psi_{00} = B^{n+l}(x_0+ix_1)^{n} (x_2+ix_3)^{l}e^{-\frac{B}{4}x^2_\mu},
\end{equation}
and the eigenfields are states of a generalized \(J_z\) operator. Orthogonality of eigenfields is given by 
\begin{equation}
\langle\psi_{n,0}| \psi_{q,0}\rangle = \int^{\infty}_{-\infty} dx_0 dx_1 (B)^{n+q} (x_0+ix_1)^q (x_0-ix_1)^n e^{-\frac{B(x^2_0+x^2_1)}{2}}
\end{equation}
which written in polar coordinates gives
\begin{equation}\label{normalized}
  \langle\psi_{n,0}| \psi_{q,0}\rangle=  2 \pi\delta_{q,n}\int^\infty_0 dr r^{n+q+1}e^{-\frac{B^2}{2}r^2} = \frac{2\pi}{B} n! (2B)^n.
\end{equation}
The normalized eigenfields are then
\begin{equation}\label{norm fields}
    \psi_{nl} = \frac{B}{2\pi}\bigg(\frac{B^{n+l}}{2^{n+l}n! l!}\bigg)^{1/2}(re^{i\theta})^n(se^{i\beta})^l e^{-\frac{B}{4}(r^2+s^2)}
\end{equation}
where \(x_0\) and \(x_1\) have been put into polar coordinates \(r\) and \(\theta\) and similarly \(x_2\) and \(x_3\) are in terms of \(s\) and \(\beta\). The expectation of \(\langle r^2 \rangle\) in the \(x_0-x_1\) plane where \(n=l=0\) is
\begin{equation}
  \langle r^2 \rangle=  \frac{2\pi}{C_0}\int^{\infty}_0 r^3 e^{-\frac{Br^2}{2}} = \frac{2}{B}
\end{equation}
If we consider some radius \(R^2>r^2\) then the number of \(n=l=0\) modes that fall in that radius, i.e. the degeneracy of the \(n=l=0\) state is \(Deg \big(\psi_{00}\big) = \frac{B^2}{4\pi^2} \beta V\) where \(V\) is the spatial volume.

\section{Field Theory Calculation}\label{sec 5}
\paragraph{}
In this section, we calculate the zero-temperature and infinite volume-pressure as \(P=\frac{\ln Z}{\beta V}\). Although we are working with a gauge theory, because the gauge is fixed, the theory is ghost-free, and Faddeev-Popov quantization is not necessary. For a complex scalar theory to calculate \(\ln Z= -\ln\det(\hat{\theta})\) we will use the identity first noted by Hawking \cite{hawking1977zeta} of
\begin{equation}
    \ln\det(\hat{\theta}) =- \frac{d}{ds}\bigg[\sum_i\frac{1}{\lambda_i^s}\bigg]_{s=0}.
\end{equation}
Using the integral definition of the gamma function, it then follows that
\begin{equation}\label{gam}
    \ln\det(\hat{\theta}) = -\frac{d}{ds}\Big[\frac{1}{\Gamma(s)} \int^\infty_0 d\tau \tau^{s-1}k(\hat{\theta})\bigg]_{s=0}
\end{equation} where \(k(\hat{\theta}) = \text{Deg}\sum_{n,l}e^{-\tau \lambda_{n,l}}\) is the heat kernel of \(\hat{\theta}\), and \(\text{Deg}\) is the degeneracy of the \(n=l=0\) state \cite{bertlmann2000anomalies}. Inserting the eigenspectrum from \eqref{spec}, the heat kernel gives a geometric series under the variable substitution of \(x= e^{-2B\tau/\mu^2}\), and evaluates to 
\begin{equation}\label{kern}
 k(\hat{\theta}) = \beta V\frac{B^2}{4 \pi^2}\sum_{n,l} e^{-B\tau[(2n +1)+(2l+1)]/\mu^2} = \beta V\frac{B^2}{16 \pi^2}\frac{1}{\sinh^2{(B \tau/ \mu^2)}}.
\end{equation}
where we have added a scale \(\mu\) to make the eigenvalues dimensionless. We will later require that \(Z\) is independent of \(\mu\) which generates the associated running coupling, and beta-function of the theory. Substituting the kernel of \eqref{kern} into \eqref{gam} yields the pressure of the system as
\begin{equation}\label{FE}
  P=  \frac{\ln Z}{\beta V} =-\frac{B^2}{e^2}+ \frac{B^2}{16 \pi^2}\frac{d}{ds}\bigg[\frac{1}{\Gamma(s)} \int^\infty_0 d\tau \tau^{s-1}\frac{1}{\sinh^2{(B \tau/ \mu^2)}}\bigg]_{s=0}.
\end{equation}
Using the identity 
\begin{equation}
    \frac{1}{\sinh^2(\tau)} = 4\sum_{n=0}^\infty n e^{-2\tau n}
\end{equation}
the integral in \eqref{FE} reduces to a gamma function after commuting the sum over \(n\) to outside of the integral giving 
\begin{equation}
    P =-\frac{B^2}{e^2} +  \frac{B^2}{4\pi^2}\frac{d}{ds}\bigg[\bigg (\frac{2B}{\mu^2}\bigg)^{-s}\zeta(s-1)   \bigg]_{s=0}
\end{equation}
which reduces to
\begin{equation}
  P= -\frac{B^2}{e^2} + \frac{B^2}{48 \pi^2}\Bigg[\ln\bigg(\frac{2B}{\mu^2}\bigg) +12 \zeta'(-1)\bigg)\Bigg].
\end{equation}
Treating \(e\) as a running coupling \(e= e(\mu)\) gives a beta-function of 
\begin{equation}
    \frac{d e(\mu)}{d\ln (\mu)} = \frac{e^3(\mu)}{48 \pi^2},
\end{equation}
matching the known one-loop beta function for scalar QED \cite{ghasemkhani2017one}. This result is consistent with the fact that we froze out all photon interactions by setting a classical background field and have included all quadratic or one-loop contributions of the theory.
In total, this gives a renormalized pressure of 
\begin{equation}
    P= \frac{B^2}{48 \pi^2}\Bigg[\ln\bigg(\frac{2B}{\Lambda^2}\bigg) +12 \zeta'(-1)\bigg)\Bigg]. 
\end{equation}
Where \(\Lambda\) is an arbitrary mass scale that must be determined empirically. Similar results are discussed in \cite{grable2023fully}. A plot of the pressure is given in figure \ref{fig 1} with \(\Lambda =1\). Beyond the minimum value, the pressure increases monotonically with \(B\).
\label{fig 1}
\begin{figure}[h]
\centering
\includegraphics[width=.8\textwidth]{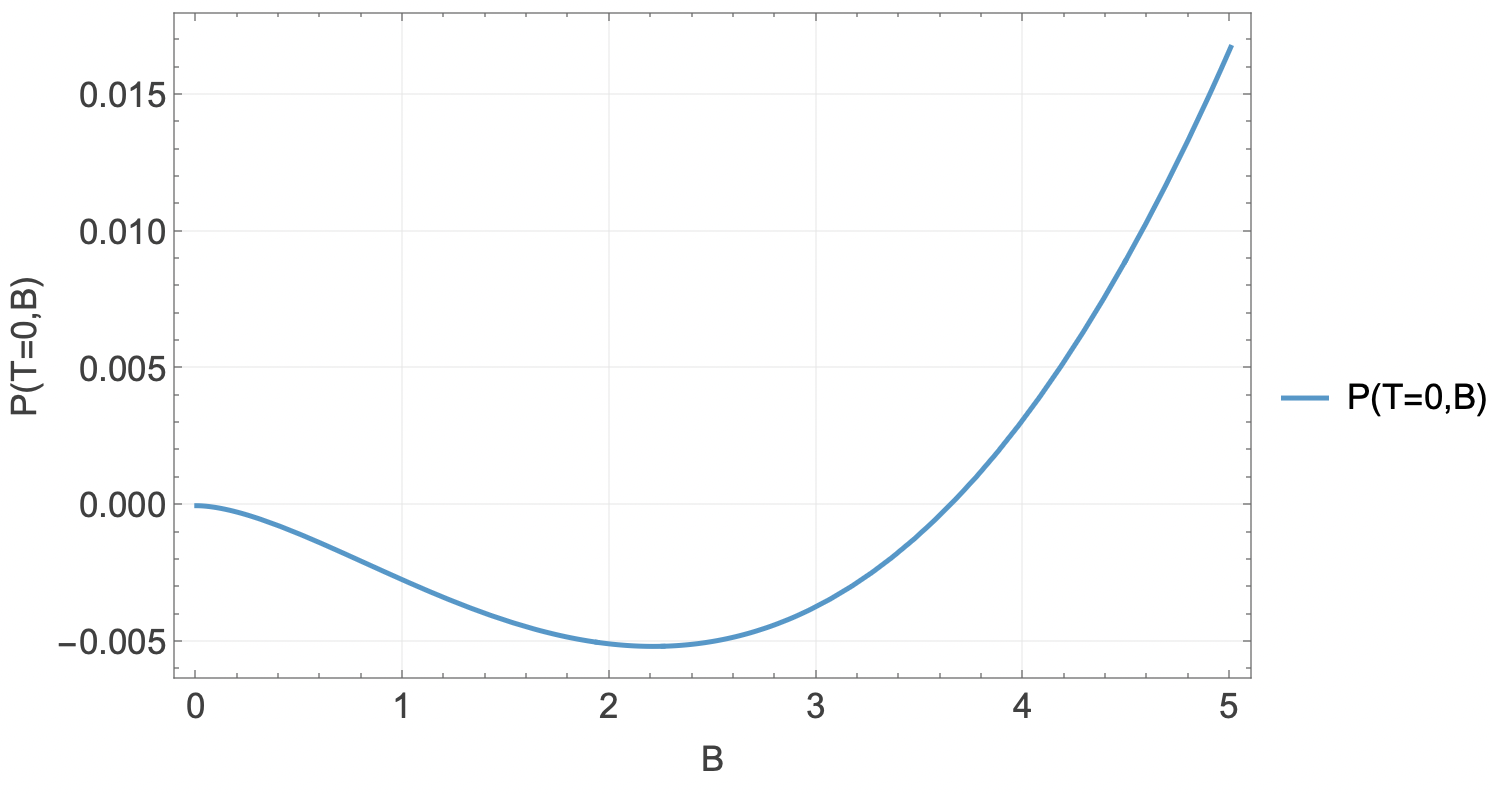}
\caption{\label{graph} P(T=0, B) is plotted with \(B\) normalized by a power of \(\Lambda^2\) such that \(B\to \Lambda^2B\) for simplicity.}
\label{runnin}
\end{figure}

\section{Calculation of the propagators}\label{Sec 6}
\paragraph{}
To calculate the propagator, we consider the two-point function as
\begin{equation}
  G(X-Y) =  \langle \phi(X)^*\phi(Y)\rangle = \frac{1}{Z} \int\mathcal{D}\phi\mathcal{D}\phi^* e^{-\int_x  \phi^*[-D^2_\mu]\phi}\phi(X)^*\phi(Y)
\end{equation}
where we have dropped the field strength tensor term, as it contributes only an overall constant to the partition function. Using the expansion of \(\phi(x,y)\) in the Landau level basis 
\begin{equation}
    \phi(x,y) = \sum_{n,l} \psi(x,y)_{nl} a_{nl}
\end{equation}
where \(\psi\) are the normalized eigenfields given in \eqref{norm fields}, the propagator \(G(0)\) is given as 
\begin{equation}
    G(0)=\langle \phi^*(x,y)\phi(x,y)\rangle = \frac{1}{Z} \int \prod d a_{nl}da^*_{nl} e^{-\sum_{n,l}2B(n+l+1)|a_{nl}|^2} \sum_{n,l}|\psi(x,y)_{nl}|^2 |a_{nl}|^2
\end{equation}
where a \(\delta\)-function contraction over the fields arises from the integral in the exponent of Z. After integrating out the expansion coefficients \(a_{nl}\) 
\begin{equation}
    G(0) = e^{-\frac{B}{2}(r^2+s^2)}\frac{B}{8\pi^2}\sum^{\infty}_{n,l=0}\frac{1}{(n+l+1)} \bigg(\frac{B}{2}\bigg)^{n+l}\frac{1}{n!l!} r^{2n}s^{2l}. 
\end{equation}
Using the identity \(\int_0^1 t^{n+l}dt = \frac{1}{(n+l+1)}\), and commuting the integral over \(t\) to outside the sum, gives
\begin{equation}\label{G(0)}
    G(0) =  e^{-\frac{B}{2}(r^2+s^2)} \frac{B}{8 \pi^2} \int_0^1 e^{\frac{B}{2}t(r^2+s^2)} = \frac{e^{-\frac{B}{4}R^2}}{2 \pi^2 R^2}\sinh\Big({\frac{B}{4}R^2}\Big)
\end{equation}
where \(R^2 = (r^2 + s^2)\). An analogous calculation yields the spatially separated propagator as
\begin{equation}\label{G(R)}
    G(R_2-R_1) =  \frac{e^{-\frac{B}{4}(R_1^2+ R^2_2-\omega)}}{2\pi^2 \omega}\sinh{\Big(\frac{B\omega}{4}\Big)}
\end{equation}
where \(\omega = r_1r_2e^{i\Delta \theta }+ s_1s_2 e^{i\Delta \beta}\).

\subsection{Propagator Limits}
\paragraph{}
The form of \(G(0)\) is naturally finite and exhibits a Gaussian profile. The interaction of the field with itself falls off as \(\frac{1}{R^2}\) from any given region of space. This is exhibited by taking the large \(R\) limit in \eqref{G(0)}, giving 
\begin{equation}\label{G(0) lim}
    \frac{1}{4\pi^2B R^2}\Big[1-e^{-\frac{B}{2}R^2}\Big]\approx \frac{1}{4\pi^2 R^2}
\end{equation}
for large \(R\). This is, however, not how two regions of the field interact over some spatial separation, and is instead the self-interactions of a single region of the field. Taking the limit that \(R\) goes zero gives
\begin{equation}
    \lim_{R\to 0} G(0) = \frac{B}{8\pi^2}
\end{equation}
\begin{figure}[h]
\centering
\includegraphics[width=.8\textwidth]{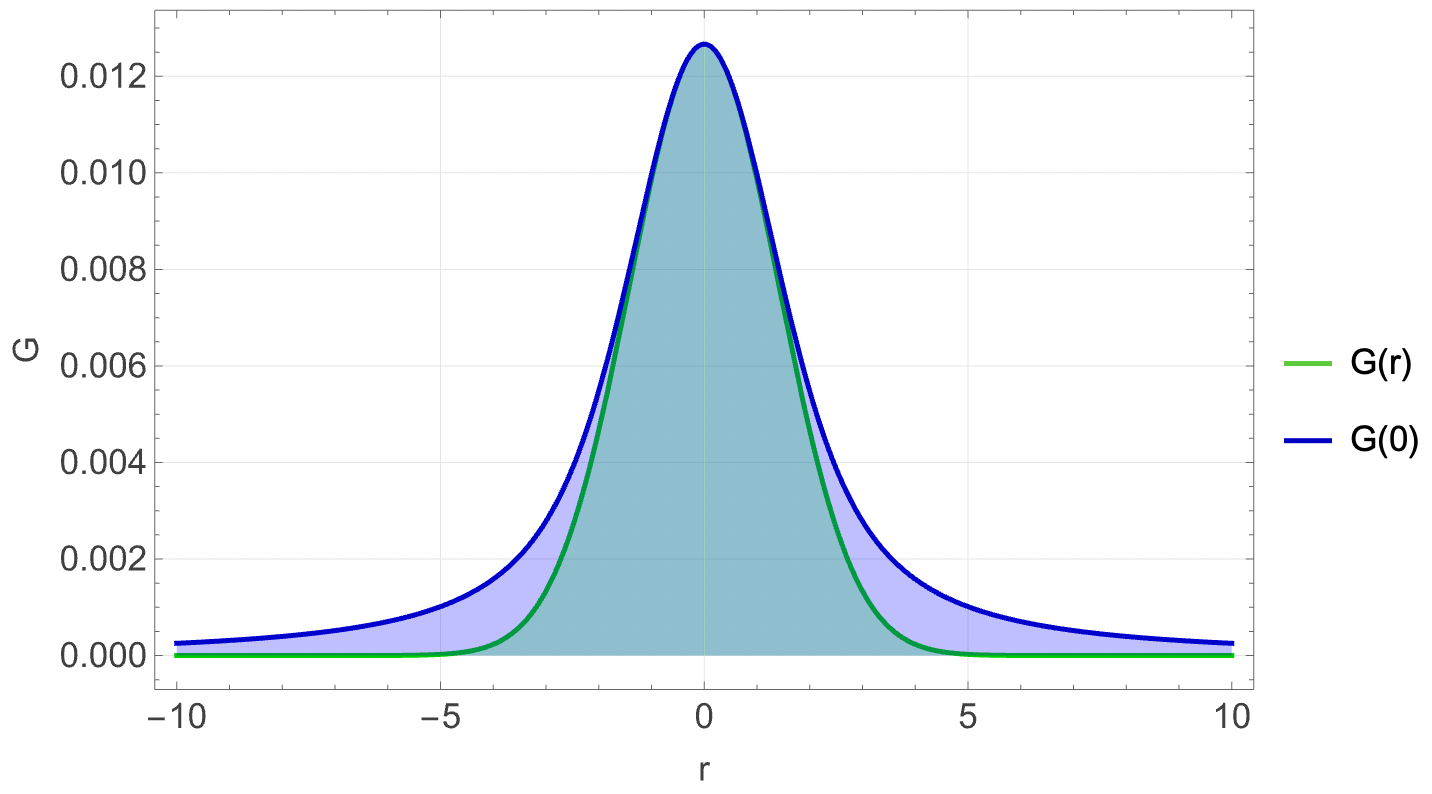}
\caption{\label{graph}G(0) and G(r) are plotted as a function of r with \(B=1\) for simplicity.}
\label{runnin}
\end{figure}
consistent with the presence of a uniform \(B\)-field throughout the vacuum. The spatial dependence in \eqref{G(0)} is an artifact of this uniform B-field providing a natural regularization of G(0) at \(R=0\). To examine how two spatially separated field regions interact, we evaluate \eqref{G(R)} with \(s_1,s_2\) and \(\Delta\theta\) all set to zero, then 
\begin{equation} \label{r1-r2 prop}
    G(r_1-r_2)\Big|_{\Delta\theta = 0} = \frac{e^{-\frac{B}{4}(r_1^2+ r^2_2-r_1r_2)}}{2\pi^2 r_1r_2}\sinh{\Big(\frac{B r_1 r_2}{4}\Big)}.
\end{equation}
Then taking the \(r_2 \to 0\) limit yields
\begin{equation}
  G(r_1) = \frac{B e^{-\frac{B}{4}r_1^2 }}{8 \pi ^2}
\end{equation}
giving a Gaussian decay across the spatial and temporal separation of the fields.
Taking the \(B\to 0\) limit in \ref{Z 2}, yields a free scalar theory with an additional surface term \(F^2_{\mu\nu}\). To take the \(B\to 0\) limit of \eqref{G(R)} we note $B$ also defines a length scale of the theory, which becomes infinite as B approaches zero. To account for the double limit, let
\begin{equation}\label{B scale def}
    B = \frac{\alpha}{R^2_2}
\end{equation}
 where $\alpha$ is a dimensionless constant. The initial coordinates \(R_1\) are a finite starting points and are therefore not included \ref{B scale def}. The substitution of \ref{B scale def} into \ref{G(R)} gives
\begin{equation}
G(R_2-R_1)  =\frac{1}{2\pi^2\omega}e^{-\frac{\alpha}{4R^2_2}(R^2_1+R^2_2-\omega)}\sinh\left(\frac{\alpha\omega}{4R^2_2}\right)
\end{equation}
The exponential dependence on $R_1$ cancels, and the remaining terms are expanded to first order to give 
\begin{equation}
\lim_{B\to0}(G(R_2-R_1)) \approx \left[\frac{1}{2\pi^2\omega}e^{-\alpha/4} \left(1 + \frac{\alpha\omega}{4R^2_2} \right)\left( \frac{\alpha\omega}{4R^2_2}\right)\right].
\end{equation}
Neglecting terms higher then \(\mathcal{O}(\frac{1}{R_2^2})\) gives
\begin{equation}
\lim_{B\xrightarrow[]{}0}(G(R_2-R_1)) \approx e^{-\alpha^2/4}\frac{\alpha}{8\pi^2R^2_2},
\end{equation}
recovering a power law decay for small \(B\) similar to the large distance limit of a free field theory propagator.

\section{Conclusion}
\paragraph{}
In this work we presented an analytically solvable model of semi-classical scalar QED in a constant self-dual background field. The self-dual configuration with constant parallel electric and magnetic fields yields closed-form propagators and an exactly solvable partition function at all field strengths.

 The matter field Hamiltonian was shown to be positive-definite, Hermitian, and gapped, with a spectrum of non-interacting bound state excitations organized as a double quantum harmonic oscillator. The energy gap \(E_{\text{gap}}=\sqrt{3B/2}\) is a direct consequence of the first quantization in the quantum harmonic oscillator and Landau level basis, and is a novel feature absent in standard translation-invariant field theories. Using zeta-function regularization and heat kernel techniques, we computed the non-perturbative partition function and extracted a beta function in agreement with the known result for scalar QED, confirming the consistency of the non-perturbative approach. We note that the theory is gapped and generates bound-states in the absence of a negative \(\beta\)-function. Finally, the matter field propagators are computed exactly in closed form in the Landau level basis, exhibiting Gaussian decay across spatial and temporal separations — a direct signature of the confining nature of the background field.

Taken together, these results demonstrate that the self-dual background generates a field theory of bound states rather than freely propagating quanta, with the constant background acting as a confining medium. The Gaussian decay of the propagators and the gapped spectrum suggest this model serves as a useful toy model for confinement-like dynamics in more complex gauge theories. Natural extensions of this work include generalizing the theory to fermionic matter fields to study the photon polarization tensor, vacuum refractive index, and birefringence relevant to magnetar phenomenology, chiral symmetry breaking, and the chiral magnetic effect. Additionally, extending the theory to non-Abelian gauge groups would allow for studies of the Savvidy vacuum and chromomagnetic flux models of the QCD vacuum.

\bibliographystyle{unsrt}
\bibliography{bibliography.bib}
\end{spacing}
\end{document}